\documentclass[aps,pra,superscriptaddress,twocolumn,amsmath,amsfonts,amssymb,floatfix]{revtex4-1}
\usepackage{enumerate}
\usepackage{mathtools}
\usepackage{amsmath}
\usepackage{graphicx}
\usepackage{epsfig}
\usepackage{sidecap}
\usepackage{hyperref}
\usepackage{color}
\usepackage{subfigure,array}
\usepackage{verbatim}
\usepackage{multirow}
\usepackage{datetime}

\usepackage{enumerate}
\usepackage{bm}

\begin{document}

\title{Exact Analytical Vortex Solution for a Two-Dimensional Quantum Gas with LHY Correction}

\author{Ibrar}\thanks{Current Address: Electrical and Electronic Metrology Division, CSIR–National Physical Laboratory, New Delhi 110012, India.}
\affiliation{Department of Physics, School of Engineering and Applied Sciences, Bennett University, Greater Noida, UP-201310, India}
\author{Mahammad Ahmed Hussain}
\affiliation{Department of Physics, School of Engineering and Applied Sciences, Bennett University, Greater Noida, UP-201310, India}


\author{Ayan Khan} \thanks{ayan.khan@bennett.edu.in}
\affiliation{Department of Physics, School of Engineering and Applied Sciences, Bennett University, Greater Noida, UP-201310, India}
\begin{abstract}
In this investigation, we provide an exact analytical vortex solution for a Bose liquid in two dimensions with beyond mean-field correction (BMF). Analytical solutions in two-dimensional systems with BMF corrections are rarely found in the literature. The present result provides a clear framework for understanding vortex structures in low-dimensional quantum fluids and serves as a reliable benchmark for future theoretical and experimental studies.
\end{abstract}

\date{\today}

\maketitle
\section{Introduction}
Physics at a lower dimension, especially in two-dimensions, is always nontrivial and exotic, the most obvious example being the $CuO_2$ superconductors \cite{bednorz1986possible}. Similarly, the Berezinskii–Kosterlitz–Thouless (BKT) transition in two dimensions is another celebrated idea which describes the phase transition of vortices from unpaired to paired states beyond a critical temperature \cite{kosterlitz1973ordering}. BKT transition was first verified in superfluid $^{3}He$ and $^{4}He$ mixture \cite{mcqueeney1984surface}. However, the direct visualization of vortex and anti-vortex pairing and unbinding has only happened in 2D ultra-cold atomic gases \cite{hadzibabic2006berezinskii}. 

In practice, ever since the experimental observation of Bose-Einstein condensate (BEC) in ultra-cold atomic gases \cite{davis, anderson1995observation}, it is considered as the biggest testing ground for condensed matter physics due to the extreme tunability and controllability of the system.

About a decade back, in two seminal papers D. S. Petrov {\it et al.} had predicted a unique liquid-like state in ultra-cold gases in three, two, and one dimension for binary BEC \cite{petrov2015quantum, petrov2016ultradilute}. The predicted liquid is qualitatively different from the usual liquid which is governed by Van-der Waal's theory.
Subsequent experimental verification of this exotic liquid in dipolar and binary BEC has firmly established the theoretical prediction \cite{kadau, pfau, che, cab}. Here, the fundamental mechanics is derived from the fact that while the many-body system is at the verge of collapse from mean-field viewpoint it is being stabilized by the beyond mean-field (BMF) interaction \cite{lee1957eigenvalues, lee1957ground}. This stabilization supports clustering of atoms to form a liquid-like state. Further, it was shown that there is a possibility of transition from droplet to a more curious state known as super-solid phase which happened to be superfluid but posses lattice like symmetry \cite{tanzi2019observation, chomaz2019long, bottcher2019transient}. Soon after this, droplets and super-solid like phases were also reported in two dimensions \cite{cattani2021two, norcia2021two}.


It must be noted that mean-field theories, such as the Gross-Pitaevskii or Bogoliubov approaches, have historically provided the cornerstone for understanding weakly interacting Bose and Fermi liquids. However, in two dimensions their predictive power is intrinsically limited by strong fluctuation effects and infrared divergences, which lead to nontrivial renormalization of basic physical quantities such as the coupling constant, the sound velocity, and the equation of state \cite{fisher2007critical, andersen2004theory}. To capture the physics of BMF contribution, it is essential to incorporate higher-order quantum fluctuations. The celebrated Lee-Huang-Yang (LHY) correction was originally derived as the first BMF term for weakly interacting Bose gasses, representing quantum depletion effects arising from zero-point motion of elementary excitations \cite{lee1957eigenvalues, lee1957ground}.


Despite substantial progress in using numerical simulations, renormalization-group methods, and perturbative expansions, fully analytical solutions incorporating the BMF contributions in two dimensions remain elusive. The difficulty stems from the logarithmic behavior of quantum fluctuations in 2D, which can invalidate naive perturbation theory unless handled with careful regularization and resummation techniques \cite{holzmann2008beyond, proukakis2008finite}. A transparent analytical theory that captures both universal fluctuation effects and system-specific properties would therefore constitute a significant advance in the understanding of low-dimensional quantum matter.

In recent years, we have observed couple of investigations in two-dimensional ultra-cold gasses through the variational method \cite{paredes2022vortex, paredes2024variational}. Specifically, in Ref.~\cite{paredes2024variational}, the variational calculation had taken into account the BMF interaction. The analysis provides an approximate variational profile that describes a vortex state. Very recently, collapse of vortices in two dimensions have been reported albeit discrepancies were noted between experiment and simulations where the possible origin is suggested as the BMF interaction effects in the fragmentation dynamics \cite{banerjee2025collapse}.

In this article, we study the two-dimensional ultra-cold gas including the BMF contribution and obtain an analytical vortex solution. We observe the variation of the vortex core as a function of topological charge $l$. To the best of our knowledge, any exact analytical solution is yet to be reported in two dimensions. We further check the stability of the solution via Vakhitov–Kolokolov (VK) criterion \cite{vakhitov1973sov}. We further calculate the vortex energy and the circulating current as functions of the topological charge and the diameter of the vortex core.

Despite considerable progress through numerical and variational methods, an exact analytical
description of vortex states in two-dimensional quantum liquids that includes BMF effects is still largely unexplored. In particular, the presence of logarithmic nonlinearity arising from LHY corrections makes analytical treatment challenging. In this work, we bridge this gap by deriving an exact analytical vortex solution and studying its stability, energy, and current density.

The report is arranged in the following way, we summarize the analytical calculation in Sec.~\ref{calculation} leading to the presentation of the vortex solution in Sec.~\ref{result}. We draw our conclusion in Sec.~\ref{conclusion}. 

\section{Calculation}\label{calculation}
In three dimensional BEC, the dynamics of the condensate is governed by the well known Gross-Pitaevskii (GP) equation where the short range two-body interaction manifests itself as a cubic nonlinear term of the condensate order parameter. The BMF interaction in three dimensions is assimilated as a repulsive quartic nonlinearity \cite{petrov2015quantum} while in lower dimensions the manifestation of the BMF interaction is relatively different. In 1D it appears as a quadratic nonlinearity and in 2D we observe a logarithmic nonlinearity \cite{petrov2016ultradilute}. It is interesting to note here that in quasi-one-dimensional (Q1D) system the nonlinearity is quartic compared to a 1D Bose gas \cite{debnath2021investigation,khan2022quantum}. In recent years, we have seen an enormous amount of interest and deep investigation in 1D and Q1D systems \cite{luo2021new,khan2022quantum}. 

However, there is a reasonable void in the analysis of 2D systems. Nevertheless we note that the equation of motion in two dimensions, for binary BEC with BMF contribution, is well captured in Ref.~\cite{li2018two} where the authors have shown the vortex formation in 2D quantum liquid numerically. Very recently we note the variational approach to study the vortex formation in 2D system \cite{paredes2024variational}. 

We start our analysis from Ref.~\cite{paredes2024variational} where the simplest form of the equation of motion of ultra dilute quantum gas in two dimensions without any loss can be described in dimensionless units as \cite{paredes2024variational}: 
\begin{equation}
i \frac{\partial \Psi}{\partial t} = - \nabla^2 \Psi + |\Psi|^2 \ln(|\Psi|^2)\, \Psi \label{egp}
\end{equation}
The mean-field wave function $\Psi$ describes the evolution of quantum states with time $t$, and $\nabla^2$ is the Laplacian operator in two dimensions comprised of $x$ and $y$ coordinates while we apply tight confinement in the $z$ direction. We all know that vortex solution is admissible in this type of setup where the central dark spot is surrounded by a radially symmetric distribution of the quantum fluid, which is coherent in nature, with phase winding around the center. The corresponding eigen states can be noted as \cite{fetter2009rotating},
\begin{equation}
\Psi(r,\Theta,t) = e^{-i \mu t} e^{i l \Theta} \psi(r),\label{ansatz1}
\end{equation}
where $\mu$ is the chemical potential, while $\Theta=\tan^{-1}(y/x)$, $r=\sqrt{x^2+y^2}$ are the polar coordinates in the $x-y$ plane. $l$ is the topological charge (integer in nature) which gives the number of windings of the phase around the central singularity. 

We consider the Laplacian in a cylindrical coordinate system with azimuthal symmetry and apply Eq.(\ref{ansatz1}) to Eq.(\ref{egp}). This allows us to write Eq.(\ref{egp}) in radial coordinate such that
\begin{eqnarray}
    \psi''+\frac{1}{r}\psi'-\frac{l^2}{r^2}\psi-\psi^3 \ln(\psi^2)+\mu\psi&=&0\label{radial_egp}
\end{eqnarray}

We assume the low energy angular momentum eigen states as, \cite{butts1999predicted, paredes2024variational}
\begin{equation}
\psi(r) = \frac{A r^l}{\sqrt{ e^{\alpha (r-r_0)} + \gamma }}=\phi(r) r^l,  \label{ansatz2}
\end{equation}
where $r_0$ is the location parameter or the shift of the distribution and $\alpha$ is the scale factor. We consider $A$, $\alpha$ and $\gamma$ as solution parameters that need to be solved.

Application of Eq.(\ref{ansatz2}) in Eq.(\ref{radial_egp}) results,
\begin{widetext}
\begin{eqnarray}
     &&   \frac{A\, r^{l-1}}{4\left(e^{\alpha (r - r_0)} + \gamma\right)^{5/2}}
    \Big[
         4r\gamma^2 \mu
        - 2 e^{\alpha (r - r_0)} \gamma \big(\alpha + 2l\alpha + r\alpha^2 - 4r\mu\big) 
         + e^{2\alpha (r - r_0)} \big( \alpha(-2 - 4l + r\alpha) + 4r\mu \big) \nonumber\\
       &&  - 4A^2 r^{1 + 2l}\big(e^{\alpha (r - r_0)} + \gamma\big)
        \ln\left(\frac{A^2 r^{2l}}{e^{\alpha (r - r_0)} + \gamma}\right)
    \Big]=0.\label{egp_expand}
\end{eqnarray}
\end{widetext}

Here, we assume that $e^{\alpha(r-r_0)}/\gamma<1$. This will ensure that the BMF interaction remains relatively weak and never explodes. We expand the logarithmic term using the Newton–Mercator series \cite{lukarevski2023series}, which is valid for small arguments.  This is justified under the condition $e^{\alpha(r-r_0)}/\gamma<1$, which ensures the convergence of the
series. The approximation leads to
\begin{eqnarray}
&&\ln\left(\frac{A^2 r^{2l}}{e^{\alpha (r - r_0)} + \gamma}\right)\nonumber
\simeq \ln(A^2) + \ln(r^{2l}) - \ln(\gamma)\nonumber\\
&&- \frac{e^{\alpha (r - r_0)}}{\gamma}
+ \frac{e^{2\alpha (r - r_0)}}{2\gamma^2}.\label{log_expand}
\end{eqnarray}

Substituting Eq.(\ref{log_expand}) into Eq.(\ref{egp_expand}) and carrying out trivial algebraic simplifications, we are now in a position to construct the coefficient equations for the exponents of $e^{\alpha(r-r_0)}$, 

\begin{eqnarray}
C_{01}&=&4\left[r\gamma^2\mu - r^{1 + 2l}\beta\gamma\ln(r^{2l})
- r^{1 + 2l}\beta\gamma\ln(\beta)\right.\nonumber\\&&\left.
+ r^{1 + 2l}\beta\gamma\ln(\gamma)\right], \label{C01}\\
C_{11} &=& \frac{1}{\gamma^2}\left[
4r^{1 + 2l}\beta\gamma^2 - 2\alpha\gamma^3 - 4l\alpha\gamma^3 
- 2r\alpha^2\gamma^3 + 8r\gamma^3\mu \nonumber \right.\nonumber\\&&\left.
\qquad - 4r^{1 + 2l}\beta\gamma^2 \ln(r^{2l})
- 4r^{1 + 2l}\beta\gamma^2 \ln(\beta)\nonumber\right.\\&&\left.
+4r^{1 + 2l}\beta\gamma^2 \ln(\gamma)
\right], \label{C11}\\
C_{21} &=& \frac{1}{\gamma^2}\left[
2r^{1 + 2l}\beta\gamma -2\alpha\gamma^2
- 4l\alpha\gamma^2 + r\alpha^2\gamma^2 + 4r\gamma^2\mu
\right].\nonumber\\\label{C21}
\end{eqnarray}
In these expressions, we have replaced $A^2$ by $\beta$ for our convenience.

We know that to satisfy the ansatz solution it is necessary that the coefficient equations must lead to zero. Hence, equating the coefficients to zero, we can obtain a further simplified set of equations, which enables us to determine the solution parameters in terms of the equation parameters. A simplified form of Eq.(\ref{C01}) leads to 
\begin{equation}
\gamma\mu - r^{2l}\beta \ln\!\left(\frac{r^{2l}\beta}{\gamma}\right) = 0. \label{coef_eq1}
\end{equation}
Similarly, using Eq.(\ref{C11}) and dividing it by \(2\gamma^2\) we obtain the following,
\begin{equation}
2r^{1 + 2l}\beta - \gamma\big[\alpha(1 + 2l) + r\alpha^2 - 4r\mu\big] - 2r^{1 + 2l}\beta \ln\!\left(\frac{r^{2l}\beta}{\gamma}\right) = 0. \label{coef_eq2}
\end{equation}
Further, from Eq.(\ref{C21}) one can construct an equation for $\beta$ such that 
\begin{equation}
\beta = \frac{\gamma\big[2\alpha(1+2l) - r\alpha^2 - 4r\mu\big]}{2r^{1+2l}}. \label{coef_eq3}
\end{equation}
Applying Eq.(\ref{coef_eq1}) in Eq.(\ref{coef_eq2}) we get
\begin{equation}
2r^{1 + 2l}\beta - \gamma\big[\alpha(1 + 2l) + r\alpha^2 -2 r\mu\big] = 0, \label{coef_eq21}
\end{equation}
here, we denote $\xi=\alpha(1 + 2l) + r\alpha^2 -2 r\mu$.
Adding Eq.(\ref{coef_eq3}) and Eq.(\ref{coef_eq21}), we get
\begin{equation}
\alpha(1 + 2l) = 2r\alpha^2 + 2r\mu. \label{coef_eq2plus3}
\end{equation}
This allows us to write $\xi=3r\alpha^2$. Hence, from Eq.(\ref{coef_eq21}) we can now determine $\beta$ as 
\begin{equation}
\beta = \frac{\gamma\xi}{2r^{1 + 2l}}.\label{beta}
\end{equation}
Applying the expression of $\beta$ in Eq.(\ref{coef_eq1}) we obtain 
\begin{equation}
\mu - \frac{\xi}{2r}\ln\!\left(\frac{\xi}{2r}\right) = 0.\label{coef_eq11}
\end{equation}

Here, we adopt a series of substitutions, such as $X=\frac{\xi}{2r}$, $y=\frac{\mu}{X}$ which allows us to write Eq.(\ref{coef_eq11}) in \textit{Lambert W} form such that $ye^y=\mu$ \cite{olver2010nist}. The appearance of the \textit{Lambert W} function reflects the nonlinear interplay between kinetic energy and BMF interactions. It naturally emerges from the transcendental structure of the governing equation and determines the characteristic length scale of the vortex.
This implies $y=W(\mu)$ leading to $X=\frac{\mu}{W(\mu)}$. Now we are in a position to express the solution parameter $\alpha$ in terms of the equation parameter, so that

\begin{equation}
\alpha^2 = \frac{2\mu}{3W(\mu)}, \qquad
\alpha = \pm\sqrt{\frac{2\mu}{3W(\mu)}}.\label{alpha}
\end{equation}

Applying Eq.(\ref{alpha}) in Eq.(\ref{beta}) we obtain a constrained equation for $\beta$ and $\gamma$.
It reads as follows 
\begin{eqnarray}
\frac{\beta}{\gamma} &=& \frac{3\alpha^2}{2r^{2l}}=\frac{\mu}{r^{2l}W(\mu)}. \label{beta_gamma}
\end{eqnarray}
The relation obtained shows that the ratio $\beta/\gamma$ depends on the
radial coordinate $r$ which contradicts our assumption that both parameters are constants. To address this, we restrict the relation to a characteristic length scale $r_0$, which represents the shift of the distribution or mean position of the
vortex density. This effectively defines the parameters on a physically relevant scale and ensures consistency of the analytical solution. Hence, we apply the constrained condition such that the ratio $\beta/\gamma$ is inversely proportional to $r_0^{2l}$. So, the modified relation will appear as 
\begin{eqnarray}
\frac{\beta}{\gamma} &=& \frac{3\alpha^2}{2r_0^{2l}}=\frac{\mu}{r_0^{2l}W(\mu)}, \label{beta_gamma}
\end{eqnarray}

Now,  to separate out the parameters $\beta$ and $\gamma$, we apply the normalization condition to the wave function.
\begin{eqnarray}
    N&=&2\pi \beta 
    \int_{0}^{\infty} 
    \frac{r^{2l+1}}{e^{\alpha (r - r_0)} + \gamma} \, dr,
    \,\,
    \text{where } \alpha > 0, \, \gamma > 0, \, l \geq 0.\nonumber\\
    &=&-\frac{2\pi\mu}{W(\mu)\alpha^2}\mathrm{Li}_2\!\left(-e^{\alpha r_0}\gamma\right)=-3\pi\mathrm{Li}_2\!\left(-e^{\alpha r_0}\gamma\right)\label{particle_no}
\end{eqnarray}
where $\mathrm{Li}_2(x)$ is the dilogarithm (or polylogarithm) function of order two which is defined as \cite{abramowitz1948handbook},
\begin{equation}
\operatorname{Li}_2(Z)=\sum_{k=1}^{\infty}\frac{Z^k}{k^2}.
\end{equation}
For small arguments, i.e., $|x|<<1$, the leading-order term dominates and the function
can be approximated as $\operatorname{Li}_2(Z)\approx z$. This allows us to write
\begin{equation}
\operatorname{Li}_2\!\big(-e^{\alpha r_0}\gamma\big)\approx -e^{\alpha r_0}\gamma.
\label{eq:Li2_approx}
\end{equation}
We are now in a position to write an expression for $\gamma$ in terms of the particle number. 
Using Eq.(\ref{particle_no}) and Eq.(\ref{eq:Li2_approx}) we determine
\begin{eqnarray}
\gamma=\frac{N}{3\pi e^{\alpha r_0}}.\label{gamma_value}.
\end{eqnarray}
This also enables us to write an explicit expression for $\beta$ (using Eq.(\ref{beta_gamma})).
We can now write the analytical vortex solution in two dimensions after employing the results obtained for $\alpha$, $\beta$ and $\gamma$. However, from Eq.(\ref{alpha}) we need to decide the sign of $\alpha$. 
The radial profile is taken in the form
\begin{equation}
\psi(r) = \frac{\sqrt{\beta}}{\sqrt{e^{\alpha (r-r_0)} + \gamma}}.
\label{eq:ansatz_alpha}
\end{equation}
As $r \to \infty$, the exponential term $e^{\alpha (r-r_0)}$ grows rapidly, so the denominator in Eq.~\eqref{eq:ansatz_alpha} increases. Therefore, $\psi(r) \longrightarrow 0 \quad \text{as} \quad r \to \infty.$
This corresponds to a localized and normalizable wave function yielding a vortex solution depending on the value of the topological charge. 

However, 
for $\alpha<0$, as $r \to \infty$, the exponential term decays such that $e^{\alpha (r-r_0)} \longrightarrow 0$ leads to a trivial constant solution $\psi(r) \longrightarrow \sqrt{\frac{\beta}{\gamma}} \neq 0$. 
So, for a non-trivial vortex solution we must choose the positive sign in Eq.~\eqref{alpha} which implies, $\alpha = \sqrt\frac{2\mu}{3\, W(\mu)}$.



For a physically acceptable localized condensate wave function, the parameters 
$\alpha$, $\beta$ and $\gamma$ must be positive, since they determine the localization
width, amplitude and finite background density of the vortex state. From the expressions derived
for $\alpha$, $\beta$ and $\gamma$ mathematical consistency requires $\frac{\mu}{W(\mu)} > 0$ or $W(\mu) > 0$ \cite{corless1996lambert}. However, the Lambert $W$ function is only real when  $\mu\geq-\frac{1}{e}$. So $\mu$ can take values greater than or equal to $\sim -0.3679$. 
In contrast, if $W(-\mu)$ is complex, $\alpha$ would become imaginary, leading to an oscillatory or divergent density distribution, which is not physically meaningful for a circulating vortex.
\section{Result}\label{result}
\begin{figure}
    \centering
    \includegraphics[width=0.8\linewidth]{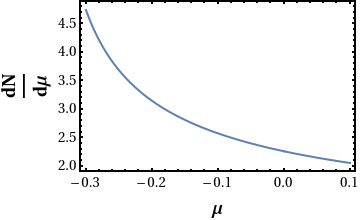}
    \caption{(Color Online) Variation of $dN/d\mu$ with chemical potential which clearly shows that $dN/d\mu>0$ suggesting stable solution.}
    \label{stability}
\end{figure}
We are now in a position to analyze the obtained solution; however, before that, it is important to check the stability of the solution. For this purpose, we employ the Vakhitov-Kolokolov (VK) criterion \cite{vakhitov1973stationary}. The VK criterion has been widely used in determining the stability of the solutions of nonlinear
Schr\"odinger equation (NLSE), which predicts the parameter regime in the chemical potential where the localized structure amplitude can grow or decay exponentially \cite{debnath2021investigation}. The VK
criterion states that a necessary stability condition is a
positive slope in dependence on the number of atoms
with respect to the chemical potential. If, $dN/d\mu>0$, the solution is
found to be stable, while $dN/d\mu<0$, the solution is unstable. The instability threshold is obtained when $dN/d\mu=0$. In Fig.~\ref{stability} we depict the variation of $dN/d\mu$ as a function of $\mu$ which clearly suggests a positive slope, thereby affirming our solution to be stable. 

\begin{figure}
\centering
\includegraphics[width=0.4\textwidth]{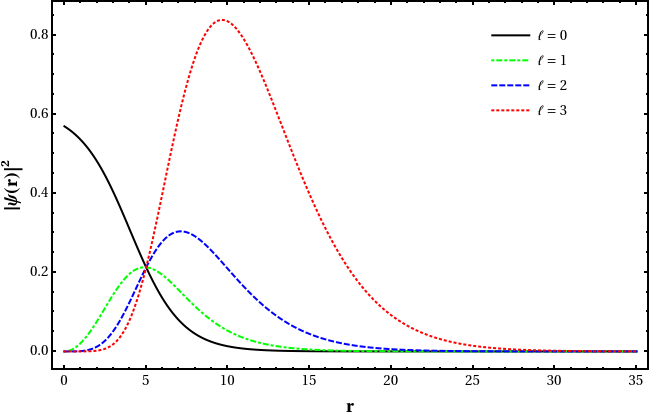}
\caption{(Color Online) The variation of the axial density for different topological charge is depicted. Here we have used $N=50$, $l=0, 1, 2, 3$, $\mu=-0.3$, $r_0=5$. }\label{radial}
\end{figure}

In Fig.~\ref{radial} we depict the radial variation of the vortex solution using Eq.(\ref{eq:ansatz_alpha}). We plotted for $N=50$ and with different values of $l$. 
The solid black line corresponds to $l=0$ while the green dot-dashed line is denoted for $l=1$. The blue dashed line depicts $l=2$ and the red dotted line describes $l=3$. The chemical potential is fixed at $-0.3$ while $r_0=5$. This clearly illustrates the motion of the peak density location with the radial coordinate. In addition, it also suggests a higher density accumulation for higher $l$ values. 

\begin{figure*}
\includegraphics[width=0.23\textwidth]{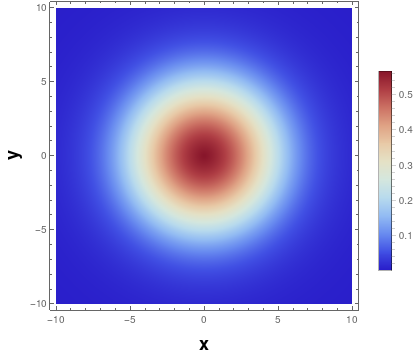}
\includegraphics[width=0.23\textwidth]{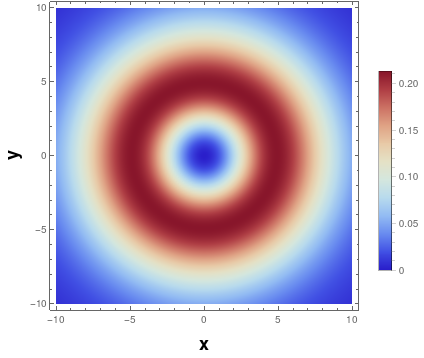}
\includegraphics[width=0.23\textwidth]{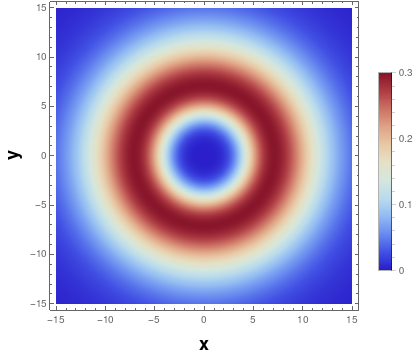}
\includegraphics[width=0.23\textwidth]{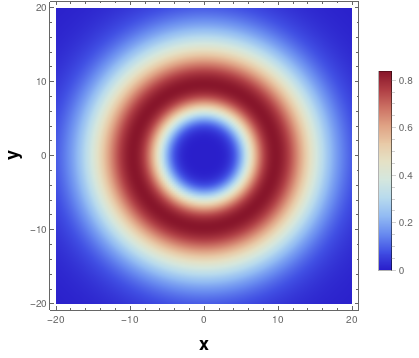}
\caption{(Color Online) Vortex generation for different topological charges. The left top corresponds to $l=0$, vortex described in the right top is for $l=1$ while the vortices in the lower panel is descibed for $l=2$ and $l=3$ (from left to right). Here, $N=50$, $\mu=-0.3$, $r_0=5$}\label{2d_density}
\end{figure*}

In Fig.~\ref{2d_density} we plot Eq.(\ref{eq:ansatz_alpha}) in the $x-y$ plane; this clearly suggests the evolution of vortex with changing values of $l$. This also shows the variation of the diameter of the vortex core. At $l=0$ there is no phase winding and thus no vortex structure. It represents the ground state of the quantum system acting as a no vortex reference against the excited vortex states. The increase in vortex core size with topological charge can be attributed to the enhanced centrifugal barrier associated with higher angular momentum states, which pushes the density outward from the center.

\begin{figure}
    \centering
    \includegraphics[width=0.8\linewidth]{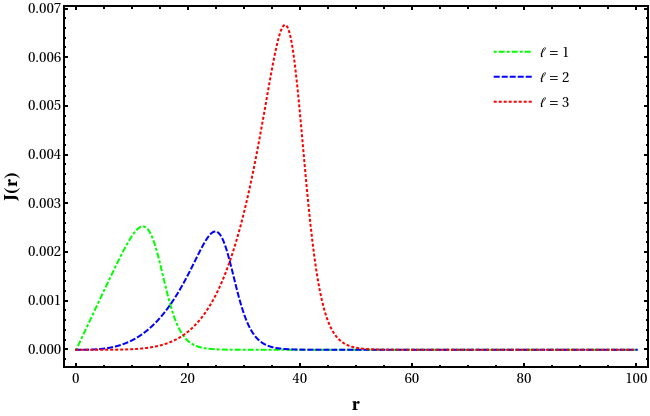}
    \caption{Variation of current density for different topological charges. Here, $N=50$, $l= 1, 2, 3$, $\mu=-0.3$, $r_0=5$.}
    \label{current}
\end{figure}

It is trivial to calculate the vortex core energy and circulating current from here on \cite{li2018two}. The vortex core energy is noted as
\begin{eqnarray}
   && \mathcal{E}=\pi\int_0^{\infty}\left|\frac{\partial\psi}{\partial\theta}\right|^2\frac{dr}{r}\nonumber\\&&=\frac{-l^2\pi\alpha^{-2l}\beta\,\Gamma(2l)\,\mathrm{Li}_{2l}(e^{-r_0\alpha}\gamma)}{\gamma};\,\,\textrm{if}\,l>0.\label{energy}
\end{eqnarray}
Similarly, the velocity of the fluid is proportional to the gradient of the phase or $v=\nabla\theta$ (in natural units). For $l=0$ the phase 
is constant, resulting in a zero velocity field, which translates to zero kinetic energy associated with that flow. The corresponding current density can be written as 
\begin{eqnarray}
    J(r)=\mathrm{velocity}\times|\mathrm{amplitude}|^2=\frac{\beta lr^{2l-1}}{e^{\alpha(r-r_0)}+\gamma}.\label{current}
\end{eqnarray}
We have depicted the variation of current density in Fig.~\ref{current}. In the figure, the green dot-dashed line is denoted for $l=1$. The blue dashed line depicts $l=2$ and the red dotted line describes $l=3$. This shows the growth in the current with increasing topological charge or in other words with higher phase winding.

\section{Conclusion}\label{conclusion}
In this article, we have presented an analytical scheme to obtain a vortex solution in an ultra-cold atomic gas with LHY correction. The obtained solution clearly manifests the vortex nature with the variation of the core radius based on the topological charge. The stability of the solution is then checked using VK criterion and is found to be stable. We further calculate the vortex energy and current density where, as expected, we find an increase in current density with an increase in $l$.

It must be noted that any analytical solution of an ultra-cold system with LHY correction is elusive till date, hence we expect our efforts to have a significant impact in understanding the dynamics of the system and serve as a reference for future studies.
\section*{Acknowledgement} AK thanks the Anusandhan National Research Foundation (ANRF), Department of Science and Technology (DST), India,
for the support provided through the project number CRG/2023/001220.

\newpage
\bibliographystyle{apsrev4-1}
\bibliography{ms_ref}

@article{luo2021new,
  title={A new form of liquid matter: Quantum droplets},
  author={Luo, Zhi-Huan and Pang, Wei and Liu, Bin and Li, Yong-Yao and Malomed, Boris A},
  journal={Frontiers of Physics},
  volume={16},
  number={3},
  pages={32201},
  year={2021},
  publisher={Springer}
}

@article{khan2022quantum,
  title={Quantum droplet in lower dimensions},
  author={Khan, Ayan and Debnath, Argha},
  journal={Frontiers in Physics},
  volume={10},
  pages={887338},
  year={2022},
  publisher={Frontiers Media SA}
}

@article{debnath2021investigation,
  title={Investigation of quantum droplets: An analytical approach},
  author={Debnath, Argha and Khan, Ayan},
  journal={Annalen der Physik},
  volume={533},
  number={3},
  pages={2000549},
  year={2021},
  publisher={Wiley Online Library}
}

@article{vakhitov1973stationary,
  title={Stationary solutions of the wave equation in the medium with nonlinearity saturation},
  author={Vakhitov, Nazib G and Kolokolov, Aleksandr A},
  journal={Radiophysics and Quantum Electronics},
  volume={16},
  number={7},
  pages={783--789},
  year={1973},
  publisher={Springer New York Consultants Bureau}
}

@book{abramowitz1948handbook,
  title={Handbook of mathematical functions with formulas, graphs, and mathematical tables},
  author={Abramowitz, Milton and Stegun, Irene A},
  volume={55},
  year={1948},
  publisher={US Government printing office}
}

@misc{lukarevski2023series,
  title={Series and products in the development of mathematics: Series and products in the development of mathematics by Ranjan Roy, Vol. 1 pp. 776,{\pounds} 69.99 (paperback), ISBN 978-1-10870-945-3; Vol. 2 pp. 476,{\pounds} 45.99 (paperback), ISBN 978-1-10870-937-8, Cambridge University Press (2021)},
  author={Lukarevski, Martin},
  year={2023},
  publisher={Taylor \& Francis}
}

@article{fetter2009rotating,
  title={Rotating trapped bose-einstein condensates},
  author={Fetter, Alexander L},
  journal={Reviews of Modern Physics},
  volume={81},
  number={2},
  pages={647--691},
  year={2009},
  publisher={APS}
}

@article{li2018two,
  title={Two-dimensional vortex quantum droplets},
  author={Li, Yongyao and Chen, Zhaopin and Luo, Zhihuan and Huang, Chunqing and Tan, Haishu and Pang, Wei and Malomed, Boris A},
  journal={Physical Review A},
  volume={98},
  number={6},
  pages={063602},
  year={2018},
  publisher={APS}
}

@article{butts1999predicted,
  title={Predicted signatures of rotating Bose--Einstein condensates},
  author={Butts, DA and Rokhsar, DS},
  journal={Nature},
  volume={397},
  number={6717},
  pages={327--329},
  year={1999},
  publisher={Nature Publishing Group UK London}
}

@article{vakhitov1973sov,
  title={Sov. Radio Phys.},
  author={Vakhitov, MG and Kolokolov, AA},
  year={1973}
}

@article{banerjee2025collapse,
  title={Collapse of a quantum vortex in an attractive two-dimensional Bose gas},
  author={Banerjee, Sambit and Zhou, Kai and Tiwari, Shiva Kant and Tamura, Hikaru and Li, Rongjie and Kevrekidis, Panayotis and Mistakidis, Simeon I and Walther, Valentin and Hung, Chen-Lung},
  journal={Physical Review Letters},
  volume={135},
  number={7},
  pages={073401},
  year={2025},
  publisher={APS}
}

@article{norcia2021two,
  title={Two-dimensional supersolidity in a dipolar quantum gas},
  author={Norcia, Matthew A and Politi, Claudia and Klaus, Lauritz and Poli, Elena and Sohmen, Maximilian and Mark, Manfred J and Bisset, Russell N and Santos, Luis and Ferlaino, Francesca},
  journal={Nature},
  volume={596},
  number={7872},
  pages={357--361},
  year={2021},
  publisher={Nature Publishing Group UK London}
}

@article{tanzi2019observation,
  title={Observation of a dipolar quantum gas with metastable supersolid properties},
  author={Tanzi, Luca and Lucioni, Eleonora and Fam{\`a}, Francesca and Catani, Jacopo and Fioretti, Andrea and Gabbanini, Carlo and Bisset, Russell N and Santos, Luis and Modugno, Giovanni},
  journal={Physical review letters},
  volume={122},
  number={13},
  pages={130405},
  year={2019},
  publisher={APS}
}

@article{chomaz2019long,
  title={Long-lived and transient supersolid behaviors in dipolar quantum gases},
  author={Chomaz, Lauriane and Petter, Daniel and Ilzh{\"o}fer, Philipp and Natale, Gabriele and Trautmann, Arno and Politi, Claudia and Durastante, Gianmaria and Van Bijnen, RMW and Patscheider, Alexander and Sohmen, Maximilian and others},
  journal={Physical Review X},
  volume={9},
  number={2},
  pages={021012},
  year={2019},
  publisher={APS}
}

@article{bottcher2019transient,
  title={Transient supersolid properties in an array of dipolar quantum droplets},
  author={B{\"o}ttcher, Fabian and Schmidt, Jan-Niklas and Wenzel, Matthias and Hertkorn, Jens and Guo, Mingyang and Langen, Tim and Pfau, Tilman},
  journal={Physical Review X},
  volume={9},
  number={1},
  pages={011051},
  year={2019},
  publisher={APS}
}

@article{che,
  title={Bright soliton to quantum droplet transition in a mixture of Bose-Einstein condensates},
  author={Cheiney, P and Cabrera, CR and Sanz, J and Naylor, B and Tanzi, L and Tarruell, L},
  journal={Physical review letters},
  volume={120},
  number={13},
  pages={135301},
  year={2018},
  publisher={APS}
}

@article{kadau,
  title={Observing the Rosensweig instability of a quantum ferrofluid},
  author={Kadau, Holger and Schmitt, Matthias and Wenzel, Matthias and Wink, Clarissa and Maier, Thomas and Ferrier-Barbut, Igor and Pfau, Tilman},
  journal={Nature},
  volume={530},
  number={7589},
  pages={194--197},
  year={2016},
  publisher={Nature Publishing Group UK London}
}

@article{pfau,
  title={Observation of quantum droplets in a strongly dipolar Bose gas},
  author={Ferrier-Barbut, Igor and Kadau, Holger and Schmitt, Matthias and Wenzel, Matthias and Pfau, Tilman},
  journal={Physical review letters},
  volume={116},
  number={21},
  pages={215301},
  year={2016},
  publisher={APS}
}

@article{cab,
  title={Quantum liquid droplets in a mixture of Bose-Einstein condensates},
  author={Cabrera, CR and Tanzi, L and Sanz, J and Naylor, B and Thomas, P and Cheiney, P and Tarruell, Leticia},
  journal={Science},
  volume={359},
  number={6373},
  pages={301--304},
  year={2018},
  publisher={American Association for the Advancement of Science}
}

@article{anderson1995observation,
  title={Observation of Bose-Einstein condensation in a dilute atomic vapor},
  author={Anderson, Mike H and Ensher, Jason R and Matthews, Michael R and Wieman, Carl E and Cornell, Eric A},
  journal={science},
  volume={269},
  number={5221},
  pages={198--201},
  year={1995},
  publisher={American Association for the Advancement of Science}
}

@article{davis,
  title={Bose-Einstein condensation in a gas of sodium atoms},
  author={Davis, Kendall B and Mewes, M-O and Andrews, Michael R and van Druten, Nicolaas J and Durfee, Dallin S and Kurn, DM and Ketterle, Wolfgang},
  journal={Physical review letters},
  volume={75},
  number={22},
  pages={3969},
  year={1995},
  publisher={APS}
}

@article{hadzibabic2006berezinskii,
  title={Berezinskii--Kosterlitz--Thouless crossover in a trapped atomic gas},
  author={Hadzibabic, Zoran and Kr{\"u}ger, Peter and Cheneau, Marc and Battelier, Baptiste and Dalibard, Jean},
  journal={Nature},
  volume={441},
  number={7097},
  pages={1118--1121},
  year={2006},
  publisher={Nature Publishing Group UK London}
}

@article{mcqueeney1984surface,
  title={Surface Superfluidity in Dilute He 4-He 3 Mixtures},
  author={McQueeney, D and Agnolet, G and Reppy, JD},
  journal={Physical review letters},
  volume={52},
  number={15},
  pages={1325},
  year={1984},
  publisher={APS}
}

@article{kosterlitz1973ordering,
  title={Ordering, metastability and phase transitions in two-dimensional systems},
  author={Kosterlitz, John Michael and Thouless, David James},
  journal={Journal of Physics C: Solid State Physics},
  volume={6},
  number={7},
  pages={1181--1203},
  year={1973}
}

@article{bednorz1986possible,
  title={Possible high T c superconductivity in the Ba- La- Cu- O system},
  author={Bednorz, J George and M{\"u}ller, K Alex},
  journal={Zeitschrift f{\"u}r physik B condensed matter},
  volume={64},
  number={2},
  pages={189--193},
  year={1986},
  publisher={Springer}
}

@book{olver2010nist,
  title={NIST handbook of mathematical functions hardback and CD-ROM},
  author={Olver, Frank WJ},
  year={2010},
  publisher={Cambridge university press}
}

@article{corless1996lambert,
  title={On the Lambert W function},
  author={Corless, Robert M and Gonnet, Gaston H and Hare, David EG and Jeffrey, David J and Knuth, Donald E},
  journal={Advances in Computational mathematics},
  volume={5},
  number={1},
  pages={329--359},
  year={1996},
  publisher={Springer}
}

@article{paredes2022vortex,
  title={On vortex and dark solitons in the cubic--quintic nonlinear Schr{\"o}dinger equation},
  author={Paredes, Angel and Salgueiro, Jos{\'e} R and Michinel, Humberto},
  journal={Physica D: Nonlinear Phenomena},
  volume={437},
  pages={133340},
  year={2022},
  publisher={Elsevier}
}

@article{paredes2024variational,
  title={Variational model for vortex quantum droplets},
  author={Paredes, Angel and Salgueiro, Jos{\'e} R and Michinel, Humberto},
  journal={Chaos, Solitons \& Fractals},
  volume={186},
  pages={115297},
  year={2024},
  publisher={Elsevier}
}

@article{andersen2004theory,
  title={Theory of the weakly interacting Bose gas},
  author={Andersen, Jens O.},
  journal={Reviews of Modern Physics},
  volume={76},
  pages={599--639},
  year={2004},
  publisher={APS}
}

@article{fisher2007critical,
  title={Critical behavior of the two‐dimensional dilute Bose gas},
  author={Fisher, Matthew P.A. and Hohenberg, Pierre C.},
  journal={Physica C: Superconductivity},
  volume={420},
  pages={59--66},
  year={2007},
  publisher={Elsevier}
}

@article{lee1957eigenvalues,
  title={Eigenvalues and eigenfunctions of a Bose system of hard spheres and its low‐temperature properties},
  author={Lee, T.D. and Huang, K. and Yang, C.N.},
  journal={Physical Review},
  volume={106},
  pages={1135--1145},
  year={1957},
  publisher={APS}
}

@article{lee1957ground,
  title={Ground state energy of a homogeneous Bose gas},
  author={Lee, T.D. and Huang, K. and Yang, C.N.},
  journal={Physical Review},
  volume={106},
  pages={1112--1117},
  year={1957},
  publisher={APS}
}

@article{petrov2015quantum,
  title={Quantum mechanical stabilization of a collapsing Bose–Bose mixture},
  author={Petrov, D.S.},
  journal={Physical Review Letters},
  volume={115},
  number={15},
  pages={155302},
  year={2015},
  publisher={APS}
}

@article{petrov2016ultradilute,
  title={Ultradilute low-dimensional liquids},
  author={Petrov, DS and Astrakharchik, GE},
  journal={Physical review letters},
  volume={117},
  number={10},
  pages={100401},
  year={2016},
  publisher={APS}
}

@article{cattani2021two,
  title={Two‐dimensional quantum droplets in ultracold atomic mixtures},
  author={Cattani, F. and Salasnich, L.},
  journal={Journal of Physics B: Atomic, Molecular and Optical Physics},
  volume={54},
  number={8},
  pages={085301},
  year={2021},
  publisher={IOP Publishing}
}

@article{holzmann2008beyond,
  title={Beyond mean‐field theory for the two‐dimensional Bose gas},
  author={Holzmann, Markus and Chevallier, D. and Krauth, W.},
  journal={Physical Review A},
  volume={81},
  number={4},
  pages={043622},
  year={2008},
  publisher={APS}
}

@article{proukakis2008finite,
  title={Finite‐temperature models of Bose–Einstein condensation},
  author={Proukakis, Nick P. and Jackson, B.},
  journal={Journal of Physics B: Atomic, Molecular and Optical Physics},
  volume={41},
  number={20},
  pages={203002},
  year={2008},
  publisher={IOP Publishing}
}

\end{document}